\documentclass[twocolumn,showpacs,preprintnumbers,amsmath,amssymb,superscriptaddress]{revtex4}

\usepackage{graphicx}
\usepackage{dcolumn}
\usepackage{bm}

\begin{document}

\preprint{APS/123-QED}

\title{Programmable trap geometries with superconducting atom chips}

\author{T. M{\"u}ller}\affiliation{Nanyang Technological University, Division of Physics and Applied Physics, 21 Nanyang Link, Singapore 637371
}
\affiliation{Centre for Quantum Technologies, National University of Singapore, 3 Science Drive 2, Singapore 117543}
\author{B. Zhang}\affiliation{Nanyang Technological University, Division of Physics and Applied Physics, 21 Nanyang Link, Singapore 637371
}
\author{R. Fermani}\affiliation{Nanyang Technological University, Division of Physics and Applied Physics, 21 Nanyang Link, Singapore 637371
}
\affiliation{Centre for Quantum Technologies, National University of Singapore, 3 Science Drive 2, Singapore 117543}
\author{K.S. Chan}\affiliation{Nanyang Technological University, Division of Physics and Applied Physics, 21 Nanyang Link, Singapore 637371
}
\author{M.J. Lim}\affiliation{Nanyang Technological University, Division of Physics and Applied Physics, 21 Nanyang Link, Singapore 637371
}
\affiliation{Department of Physics, Rowan University, 201 Mullica Hill Road, Glassboro, NJ 08028, USA}
\author{R. Dumke}
\affiliation{Nanyang Technological University, Division of Physics and Applied Physics, 21 Nanyang Link, Singapore 637371
}

\date{\today}

\begin{abstract}
We employ the hysteretic behavior of a superconducting thin film in the remanent state to generate different traps and flexible magnetic potentials for ultra-cold atoms. The trap geometry can be programmed by externally applied fields. This new approach for atom-optics is demonstrated by three different trap types realized on a single micro-structure: a Z-type trap, a double trap and a bias field free trap. Our studies show that superconductors in the remanent state provide a new versatile platform for atom-optics and applications in ultra-cold quantum gases.\end{abstract}


\maketitle
The use of superconductors in atom chips \cite{Hinds99,Folman02,Fortagh07} is a recent development, presenting new opportunities for atom optics \cite{Nirrengarten06,Mukai07,Cano08,Roux08,Mueller08}. One demonstrated advantage of superconductors over conventional conductors is the significant reduction of near-field noise in current-carrying structures leading to low atomic heating rates and enhanced spin-flip lifetimes \cite{Emmert09,Hufnagel09,Kasch09,Scheel05,Skagerstam06,Fermani09}. Proposals in this area advocate experimental designs for coherent coupling with atomic or molecular quantum systems that exploit the distinct properties of superconductors \cite{Tian04,Sorensen04,Andre06,Rabl06,Petrosyan08,Tordrup08,Imamoglu09,Verdu09}.

In a previous paper we have demonstrated that the remanent magnetization created by vortices can be used to trap ultra cold atoms without applying a transport current \cite{Mueller09}. Other groups have created quadrupole type traps \cite{Shimizu09} and have shown that vortices modify the trapping potential created by a transport current in a Z-type trap \cite{Emmert09b}. In this article, we show that the unique response of superconductors to applied magnetic field enables programmable magnetic trap geometries for ultra cold atoms.
We demonstrate this new approach by generating three different atom trap geometries on a fixed superconducting micro-structure. We can choose the geometry by applying a suitable external magnetic field sequence. The trapping potentials are generated by spatial magnetic patterns imprinted on a thin film, using the hysteretic response of type-II superconductors.  

\begin{figure}[b!]
\includegraphics[width=8.4cm]{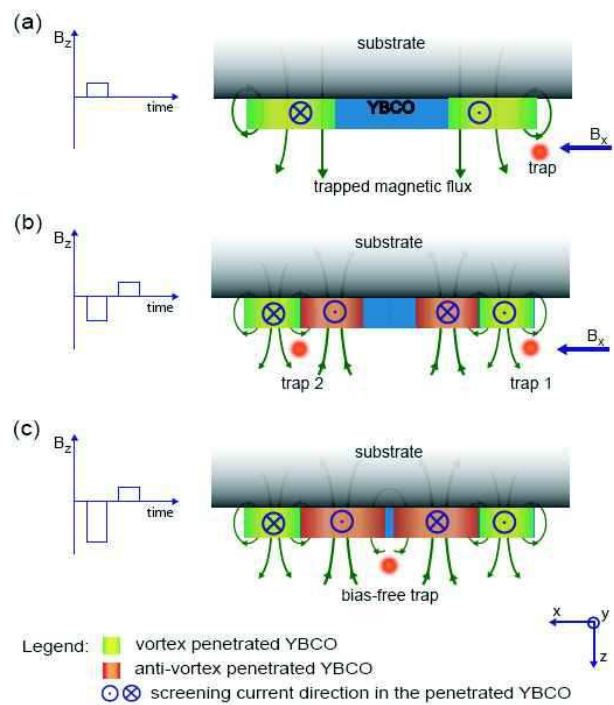}
\caption{\label{fig:1}(color online) Sequences of applied magnetic field $B_z$ imprinting patterns of magnetization on the film (left) and simplified schematics of vortex-based trap geometries (right). (a) Single magnetic trap due to the combination of the trapped magnetic flux with a magnetic bias field $B_x$.
(b) Double magnetic trap due to the combination of trapped magnetic flux and anti-flux with a magnetic bias field $B_x$.
(c) Single magnetic trap generated only by imprinted magnetic pattern, no bias magnetic field applied. 
}\end{figure}

The three different geometries we realize in this article to demonstrate this new approach are shown schematically in Fig.~\ref{fig:1}. We first study the time dependence of the single Z-type trap shown in Fig.\ref{fig:1}(a), which verifies the stability of the remanent magnetization of the film. Then, we use a distance measurement of the trapped atoms to study the hysteretic magnetization of the film. In this measurement regions of opposing magnetization are imprinted in the film, which are used to create a double trap by superimposing a bias field as represented in Fig.~\ref{fig:1}(b). Finally, we realize a bias-free trap as in Fig.~\ref{fig:1}(c) solely by imprinting a suitable magnetic field pattern on the superconducting strip. This enables us to realize trapping potentials relying only on the remanent magnetization, decoupled from external noise sources. All of our experimental findings are supported by simulations based on mesoscopic models \cite{Schuster94,Brandt96,Dikovsky09}, where the remanent magnetization is modeled by current distribution as indicated simplified in Fig.~\ref{fig:1}.\\
Our measurements are performed with the following experimental setup and procedure. We prepare laser cooled and trapped atoms in a ultrahigh vacuum chamber using standard techniques. The superconducting chip is mounted facing downward on an isolated cold-finger and cooled with liquid nitrogen to about 83K. The superconductor is a YBCO thin film with a thickness of $0.8\mu m$, which is structured using optical lithography. The vortex-based trap is realized below a $400\mu m$ wide strip, which is structured in a Z-shape with a crossing $200\mu m$ wide compression wire in its center.

Our experimental procedure starts with loading a magneto-optical trap (MOT) placed 35mm below the chip position. Afterward we apply molasses cooling and optical pumping. The atoms are then magnetically caught by fast ramping up a quadrupole field generated by the MOT coils. For an efficient transfer to the magnetic trap, a minimum field gradient of $\frac{\partial B_{z}}{\partial z}=22$G/cm at the end of the fast ramp is required to counter gravity and the motion of the atoms. At the position of the chip this gradient field has a magnetic field component $B_z=76$G. This value is higher than the first critical field of our superconductor at 83K and leads to vortex penetration. The gradient field for the magnetic catch is always present in each run of the experiment.

The magnetic catch is followed by a magnetic transfer of the atoms to a second quadrupole field centered close to the chip. The absolute magnetic field at the chip is reduced below the first critical field during this transfer. A stable distribution of vortices remains in the thin film. This leads to a macroscopic magnetization of the superconductor. As described in \cite{Mueller09}, we load the atoms in the vortex-based micro-trap by quickly switching off the second quadrupole field and turning on the homogeneous bias field. The trapping potential is generated by the combination of the bias field $B_x$ with the field of the vortices, as schematically sketched in Fig.~\ref{fig:1}(a). We load up to $10^6$ atoms in the Z-type micro-trap which has a lifetime of a few seconds.\\
\begin{figure}[h!]
\vspace{-0.5cm}
\includegraphics[width=8.4cm]{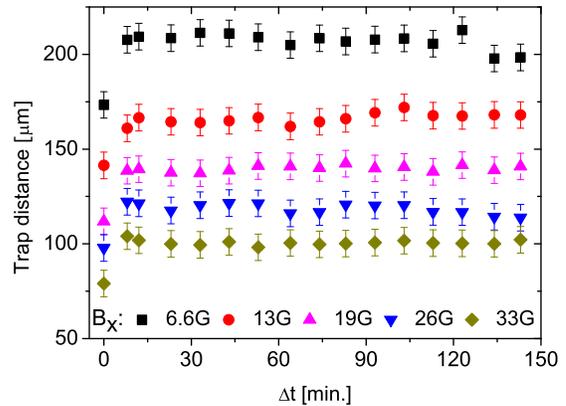}
\vspace{-0.5cm}
\caption{\label{fig:2}(color online) Stability of magnetization in the superconducting film. The trap distance for various bias fields $B_x$ is measured over time $\Delta t$ at intervals of 10 minutes. For the first measurement at $t=0$ a magnetic field of 76G is used to magnetize the film. This measurement serves as a reference for the following. Afterward an additional magnetic pulse with a field of 220G has been used once to increase the average magnetization. All subsequent measurements are therefore sensitive to vortex relaxation, which is not observed on this timescale.}
\end{figure}
We measure the distance of the surface to the atom cloud held in the Z-type trap for various bias field $B_x$ at regular time intervals over more than 140 minutes as shown in Fig.~\ref{fig:2}. The trap distance is given by the macroscopic magnetic field due to the trapped vortices \cite{Mueller09} and detect no reduction of this distance. We infer that the prepared vortices in the YBCO film remain long enough for all experiments presented in this article - which have a time scales on the order of seconds. This study verifies the long term stability of the macroscopic magnetization of the film. Our results are in agreement with with measurements of vortex relaxation in thin films by other techniques \cite{Yeshurun96}.

\begin{figure}[h!]
\includegraphics[width=8.4cm]{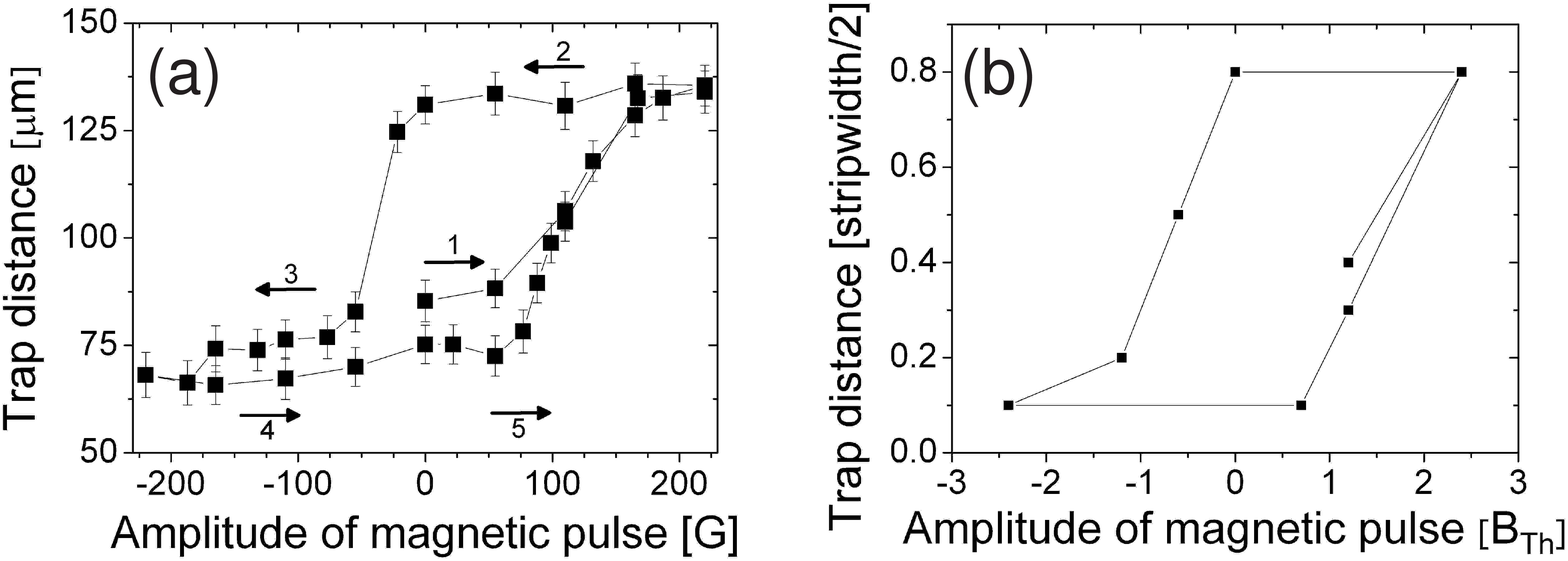}
\caption{\label{fig:3}Hysteresis of the superconducting film with respect to applied magnetic field.  (a) The trap distance is an indirect probe of the average vortex density, i.e. film magnetization. The applied field pulses were varied in the following order: 1. 0G $\rightarrow$ 220G, 2. 220G $\rightarrow$ 0G, 3. 0G $\rightarrow$ -220G (reversed), 4. -220G $\rightarrow$ 0G, 5. 0G $\rightarrow$ 220G. (b) Simulated trap distance, normalized to half the strip width ($=200\mu m$ for our film) and the thermodynamic critical field $B_{Th}$, for a bias field $B_x=0.17B_{Th}$. The shown simulation is obtained without free parameters by the mesoscopic model derived from \cite{Schuster94,Brandt96}. In the regions of reversed fields (3)-(4) a double trap is obtained, as in Fig.~\ref{fig:4}. The measured and simulated distance always refers to the (single) trap created on the same side of the micro-structure.}
\end{figure}

We next investigate the hysteretic behavior of type-II superconductors with respect to applied magnetic fields. In particular we study a sequence field reversals with a measurement of the trap position, shown in Fig.~\ref{fig:3}. The results are used to verify our simulations. The numerical simulations allow us to shape new magnetic potentials based on the remanent state in the superconductor. The measurement is carried out in a temporal sequence with 5 stages as indicated in Fig.~\ref{fig:3}(a). To study the hysteresis of the superconductor we apply additional imprinting field pulses in between each standard atom trapping procedure. These imprinting fields are changed step wise and the distance is measured after each step. However, the field of 76G during the magnetic catch is present in each measurement and has a significant influence on the remanent magnetization, in particular for reversed field pulses.

In the first part of the sequence we start from the virgin state of the superconductor. We apply increasing additional imprinting field pulses up to +220G, as indicated in Fig.~\ref{fig:3}(a) (1). For increasing field pulse amplitudes the trap center is shifted further away from the chip as expected. Afterward, we lower the imprinting field pulses, which causes no change in the distance (2). During stage (3) we reverse the orientation of the additional field pulse. This leads to anti-vortices introduced into the superconducting film and the trap position shifts towards the chip due to a reduced magnitude of magnetization in the remanent state. For increasing reversed pulses the trap center shifts closer to the chip (3). This effect continues for increasing reversed field amplitudes up to a certain characteristic value $B_R$ while for higher reversed the polarity of the magnetization can be reversed completely. This characteristic field $B_R$ depends on the thermodynamic critical field $B_{Th}$ and the formerly applied positive field (+220G in the described measurement). In the case of completely removed positive magnetization we can still realize trapping of atoms. This is due to the field of +76G present in the magnetic catch, which causes a partial positive magnetization again. However, the change in the trapping characteristics causes a less pronounced shift of the trap center, visible for reversed field pulses higher than -80G. When we apply imprinting pulses changing back from reversed (4) to positive we do not notice any change in the trap distance until the imprinting field is in the original direction and higher than 76G (5). This offset is due to the field of the magnetic catch. We increase the additional field pulses up to +220G and observe that the reversed magnetization can be completely overwritten with positive magnaetization. When repeating the described cycle a second time the trap position follows the points as measured before (not shown in Fig.~\ref{fig:3}).

We simulate the magnetic history as described in the above procedure using mesoscopic models \cite{Schuster94,Brandt96,Dikovsky09}. We use the simulations to find the trap distance dependence on the magnetic history. These results displayed in Fig.\ref{fig:3}(b) are in qualitative agreement with our measurements and are used for empirical adjustment of our model. This allows to employ the simulations as a design tool for new trapping potentials.\\
\begin{figure}
\includegraphics[width=8.4cm]{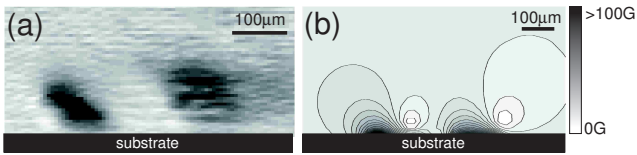}
\caption{\label{fig:4}(a)Absorption image of the atoms in a double trap which is realized on a single superconducting strip using a magnetic field sequence with pulses of +220G,-220G,76G. (b) Simulated magnetic field distribution created by spatially patterned regions of vortices and anti-vortices in the film combined with the bias field $B_x$. This creates a double trap.}
\end{figure}
The spatial magnetization patterns imprinted in the film by field reversals can be used to design complex potentials for controlled manipulation of cold atoms. We demonstrate this by realizing a double trap on the single film strip, as schematically shown in Fig.\ref{fig:1}(b). This double trap is prepared by first applying a reversed field pulse higher than -120G, followed by the field of 76G during the magnetic catch. This corresponds parts of the stages (3)-(4) in Fig.~\ref{fig:3}. The second trap is more pronounced for higher reversed fields. Shown in Fig.~\ref{fig:4}(a) is an \textit{in-situ} absorption image of the atoms stored in the double trap created by the sequence +220G, -220G, +76G. We emphasize that for this double trap the last two fields of (-200G, +76G) are the essential components. The magnetization created by the field pulse of +220G is completely removed by the reversed field and is only applied as part of the hysteresis measurement. The creation of the second trap is given by regions with opposing magnetization in the strip. This leads to a complex spatial field distribution Fig.\ref{fig:1}(b), allowing for trapping at different locations in combination with the bias field $B_x$. The simulation of this field for our parameters is shown in Fig.\ref{fig:4}(b).

\begin{figure}
\includegraphics[width=8.4cm]{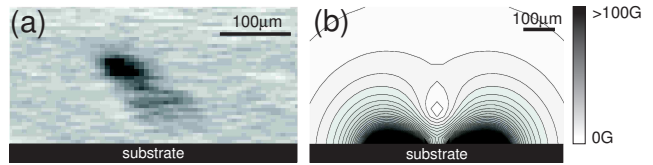}
\caption{\label{fig:5} (a) Absorption image of atoms in a magnetic micro-trap without additional bias field. The trap is created by a suitable pattern of opposing magnetization due to vortices and antivortices in the strip. (b) Simulation of the magnetic field distribution of the bias field free trap.}
\end{figure}

In the simulations we find that the applied magnetic pulses can be arranged to create a trapping potential without bias fields by the combination of the magnetic fields of vortices and anti-vortices alone. The creation of the (radial) trapping potential can be schematically described in the simplified mesoscopic model as similar to a four wire guide \cite{Dekker00}, see Fig.\ref{fig:1}(c). For our parameters this trap can be created by applying a magnetic pulse sequence of -220,+76G. The bias free trap could also be obtained with opposite signs during this sequence, i.e. +220,-76G. However, for technical simplicity we choose to employ the field of the magnetic catch as the second field of the sequence. An \textit{in-situ} absorption image of the atoms in this trap is shown in Fig.~\ref{fig:5}(a). The bias field free trap has a lifetime of about 40ms. By employing specially designed micro-structures \cite{Fernholz08} more optimized trapping geometries with longer lifetimes are in reach. The realized trap demonstrates the unique advantages provided by vortices and anti-vortices in remanent state superconductors for dynamic tuning of the trap parameters and even the trap geometry itself.

The programmable magnetic potentials created with remanent state superconductors in this article will enable new applications for atom optical manipulations of ultra-cold atoms and quantum gases. These proposed novel trapping schemes are of particular interest due to the potential low noise provided by the combination of induced magnetization and the reduced near-field noise of the superconducting material \cite{Fermani09,Scheel05,Skagerstam06}. One possible application is guiding in double structures which is of great interest for integrated atom interferometers. New approaches for atom guides based on the bias free trapping potentials are particularly interesting due to the low technical noise. This could be exploited to realize ring-shaped traps on superconducting disks or rings. With multiple appropriately adjusted field pulse reversals, periodic magnetic potentials similar to a lattice geometry could be imprinted in a superconductor. To create these potentials, the applied sequence must consist of fields with opposing polarity and decreasing magnitude, to ensure that previously prepared magnetization is not completely removed from the strip by subsequently applied fields.

In conclusion, we have shown that the hysteresis of a type-II superconductor can be used to imprint complex spatial magnetic field patterns for manipulation of ultra-cold atoms. This unique approach allows the realization of multiple trap geometries using the same micro-structured thin film, in turn enabling new experiments in atom-optics. In addition, the vortex based micro-trap is a useful tool to investigate the distribution of remanent magnetization in superconductors. The presented programmable trap geometries are of high interest for atom optical experiments, in particular with the low noise environment and bias field free trapping.\\
We acknowledge financial support from Nanyang Technological University (grant no. WBS M58110036), A-Star (grant no. SERC 072 101 0035 and WBS R-144-000-189-305) and the Centre
for Quantum Technologies, Singapore. MJL acknowledges travel support from NSF PHY-0613659 and the Rowan University NSFG program.

\end{document}